\documentclass[sigconf,10pt]{acmart}
\renewcommand\footnotetextcopyrightpermission[1]{} 
\usepackage{amsmath}
\usepackage{amssymb}
\usepackage{amsfonts}
\usepackage{array}
\usepackage{booktabs} 
\usepackage{times}
\usepackage{balance}
\usepackage[latin1]{inputenc}
\usepackage{amsmath}
\usepackage{amsfonts}
\usepackage{amssymb}
\usepackage{verbatim}
\usepackage{graphicx}
\usepackage{hyperref}
\usepackage{subfigure}
\usepackage{enumitem}
\usepackage{bm}
\usepackage{url}
\usepackage{bbm}
\usepackage{balance}	
\usepackage{epsfig}
\usepackage{mathrsfs}
\usepackage{float}
\usepackage{algorithm}
\usepackage{algpseudocode}
\algnotext{EndFor}
\algnotext{EndIf}
\algnotext{EndWhile}
\algnewcommand\algorithmicforeach{\textbf{for each}}
\algdef{S}[FOR]{ForEach}[1]{\algorithmicforeach\ #1\ \algorithmicdo}
\newcommand{\tup}[1]{\left< #1 \right>}			
\newcommand{\ind}{1}
 \setcopyright{none}
 
\begin{document}
\setlength{\abovedisplayskip}{5pt}
\setlength{\belowdisplayskip}{5pt}
\setlength{\abovecaptionskip}{0pt}
\setlength{\belowcaptionskip}{0pt}
\setlength{\textfloatsep}{5pt plus 1.0pt minus 1.0pt}


\title[Modeling Property Graphs]{When Labels Fall Short: Property Graph Simulation via Blending of Network Structure and Vertex Attributes }

\author{Arun V. Sathanur}
\affiliation{%
  \institution{Pacific Northwest National Laboratory, Seattle, WA, USA}
}
\email{arun.sathanur@pnnl.gov}

\author{Sutanay Choudhury}
\affiliation{%
  \institution{Pacific Northwest National Laboratory, Richland, WA, USA}
}
\email{sutanay.choudhury@pnnl.gov}

\author{Cliff Joslyn}
\affiliation{%
  \institution{Pacific Northwest National Laboratory, Seattle, WA, USA}
}
\email{cliff.joslyn@pnnl.gov}

\author{Sumit Purohit}
\affiliation{%
  \institution{Pacific Northwest National Laboratory, Richland, WA, USA}
}
\email{sumit.purohit@pnnl.gov}

\begin{abstract}
Property graphs can be used to represent heterogeneous networks with labeled (attributed) vertices and edges. Given a property graph, simulating another graph with same or greater size with the same statistical properties with respect to the labels and connectivity is critical for privacy preservation and benchmarking purposes.  In this work we tackle the problem of capturing the statistical dependence of the edge connectivity on the vertex labels and using the same distribution to regenerate property graphs of the same or expanded size in a scalable manner. However, accurate simulation becomes a challenge when the attributes do not completely explain the network structure.  We propose the Property Graph Model (PGM) approach that uses a label augmentation strategy to mitigate the problem and preserve the vertex label and the edge connectivity distributions as well as their correlation, while also replicating the degree distribution.  Our proposed algorithm is scalable with a linear complexity in the number of edges in the target graph. We illustrate the efficacy of the PGM approach in regenerating and expanding the datasets by leveraging two distinct illustrations. Our open-source implementation is available on GitHub \footnote{https://github.com/propgraph/pgm}.
\end{abstract}


\maketitle

\section{Introduction}
\label{sec:intro}

Most real-world datasets that naturally lend themselves to a graph representation also contain significant amounts of label (or attribute) information. This situation is promoting the popularity of property graphs: multi-graphs where the vertices and edges are labeled with key-value pairs \cite{Simeonovski:2017}.  For example, the Microsoft Academic Graph has labels such as affiliation, field of study, etc., for every person.  These attributes help answer questions such as: 1) how strong are collaborations between two fields?  2) where is a person with a certain affiliation and field of study likely to publish most?  Similar motivating examples are abound in other domains such as bioinformatics (protein-interaction networks), medicine (clinical records) and cyber-security (network-traffic data).  The need for accurate and scalable simulation arises as an important capability for property graphs. We often need to re-generate datasets with equivalent properties for privacy reasons, or expand a dataset by multiple orders of magnitude for benchmarking studies.

\textsc {\textbf{The Problem}} In this work we consider the problem of capturing the relationships between given and (in general) correlated vertex labels and edge connectivity in property graphs through the use of two different joint distributions. We show that a straightforward approach to capturing the label-structure relationships can be accuracy-limited when the given labels cannot explain the structure completely. We mitigate this problem by modeling the dependence of the edge connectivity on the vertex labels and the structure itself via the introduction of an augmented label that categorizes the vertex degree distribution. We demonstrate the modeling of graphs with vertices of the same type, connected by one specific type of relationship. General property graphs with heterogeneous vertices and multiple relationships can be modeled by introducing vertex types as new labels and building multiple distributions for the typed edges.

\textsc {\textbf{Contributions}} Our Property Graph Model (PGM) retains the \textit{generative} nature of the Multiplicative Attribute Graph (MAG) model \cite{kim2012multiplicative} by expressing the probability of edge connection as a function of the vertex labels. However, while MAG deals with latent labels, PGM caters to correlated, meaningful real-world labels. In this way it is similar to the Attribute Graph Model (AGM) approach \cite{pfeiffer2014attributed}. PGM has the added benefit of not needing to assume a model for the graph topology, making it general enough to model property graphs across domains. The use of label and edge categories to define multinomial distributions provides for a scalable implementation linear in the number of edges required in the target dataset. Finally, we demonstrate our results on two datasets: a synthetic dataset generated by a role-based approach \cite{henderson2012rolx} and a real-world dataset extracted from the Facebook Social Graph \cite{snapnets}.

\section{The Basic PGM Approach}
\label{sec:approach}
Consider a source property graph $G_S = \tup{V_S, E_S, L, L(V_S)}$ with the set of vertices $V_S$ and the set of edges $E_S \subseteq V_S \times V_S$. $L = \{ L_k \}_{k=1}^M$ is a set of $M$ vertex label sets. Associated with the $k^{th}$ label is $L_k$, the set of possible label values for that label and $n_k = |L_k|$ . For example, in a social graph, the first label, \textit{Income-Range}, may have 6 possible values where as the second label, \textit{Education-Level}, may have 4 possible values. Associated with each vertex $v_i \in V_S$ is a random $M$-vector $\bar{L}(v_i) = \tup{l^i_1,l^i_2, \ldots, l^i_k, \ldots, l^i_M}$ drawing a label value $l^i_k \in L_k$, for each of the $M$ labels. We denote by $L(V_S)$, the set of all the $|V_S|$ label value vectors in one to one correspondence with the set of vertices $V_S$.The target property graph $G_T = \tup{V_T, E_T, L,L(V_T)}$ is defined analogously, and is generated by capturing appropriate statistics on $G_S$. Note that both $G_S$ and $G_T$ share the same set of vertex labels $L$.

Each realized vertex label vector $\bar{L}(v_i)$ can be considered as a draw from the set of joint label assignments ${\mathcal L} = \bigtimes_{k=1}^M L_k$. There are $N=\prod_{k=1}^M n_k$ possible joint label assignments called \textit{label categories} and the $j^{th}$ label category is denoted by $c_j$. In doing so, we flatten the joint distribution to a multinomial label distribution $P_L$ over these $N$ categories such that $p_j = P_L\left(c_j\right)$, and $\sum_{j=1}^{N} p_j = 1$. With the observations of the vertex labels in the source dataset $G_S$, we can estimate the parameters $p_j$ via the maximum-likelihood method as
\begin{equation}
P_L\left(c_j\right) = \frac{\sum_{i=1}^{|V_S|}\ind_{c_j}(\bar{L}(v_i))}{|V_S|}.
\end{equation}
Here the indicator function is $1$ only when the label vector for vertex $i$ is equal to the joint label category $c_j$.

Next, we model the edge connectivity by a joint distribution over pairs of label categories $\left(c_j,c_{j'}\right)$ which we call {\em edge categories}. We denote {\em this} distribution by $P_C$. $P_C$ is defined over the sample space ${\mathcal L} \times {\mathcal L}$ and has one entry per pair of label vector realizations. $P_C$ can be estimated from data as
\begin{equation}
P_C\left(\left(c_j,c_{j'}\right)\right) = \frac{\sum_{\tup{v_i,v_{i'}} \in E_S}\ind_{(c_j,c_{j'})}\left(\bar{L}(v_i),\bar{L}(v_{i'})	\right)}{|E_S|}.
\label{eqn:edgecon}
\end{equation}
Here the indicator function is $1$ only when the two vertices $v_i$ and $v_{i'}$ have an edge between them and their label vectors take on categories $c_j$ and $c_{j'}$ respectively. Note that for undirected graphs, where the order of $c_j$ and $c_{j'}$ does not matter, $P_C$ can be represented as a more compact multinomial distribution with $\hat{N} =  { N \choose 2}$ categories. When we draw an edge from $P_C$, the successful category gives the vertex label categories corresponding to the two end points that form the edge. Using a data structure such as a map ($C2V$ in Algorithm \ref{algo-pgmbasic}), the participating vertices can be randomly drawn from the pools of vertices corresponding to those label categories. Drawing the edges from the multinomial distribution $P_C$ renders the algorithm linear in the number of edges required as opposed to a naive implementation over vertex pairs which will lead to an algorithm that is quadratic in the number of vertices required. Algorithm \ref{algo-pgmbasic} describes the basic PGM approach.

\begin{algorithm}[!htbp]
\begin{algorithmic}[1]
\Procedure{PGM-BASIC}{$D_S,n_t,m_t$}
\State $\tup{V_S,E_S,L,L(V_S)}$ = $processSourceDataSet\left(D_S\right)$
\State $G_T$ = \Call{simAttrGraph}{$\tup{V_S,E_S,L,L(V_S)},n_t,m_t$}
\EndProcedure
\Procedure{simAttrGraph}{$\tup{V,E,L,L(V)},n_t,m_t$}
\State $P_L$ = $computeVertexLabelDist\left(V,L(V)\right)$
\State $P_C$ = $computeEdgeConnectivityDist\left(V,E,L(V)\right)$
\State $V_X$ = $\phi$, $L(V_X)$ = $\phi$,  $E_X$ = $\phi$
\For{$idx$ = $1$ to $n_t$}
\State $\left(v,\bar{L}(v)\right) = sampleFromMultiNomialDist\left(P_L\right)$
\State $V_X = V_X \cup \left\{v\right\}$, $L(V_X) = L(V_X) \cup \left\{\bar{L}(v)\right\}$
\EndFor
\For {$i$ = $1$ to $N$} \Comment{Create map $C2V$ for all $N$ categories}
\State $C2V\left[c_i\right]$ = Set of all vertices with label category $c_i$
\EndFor
\For{$idx$ = $1$ to $m_t$} \Comment{Draw one edge at a time}
\State $\left[c_1,c_2\right] = sampleFromMultiNomialDist(P_C)$
\State Draw $v_1$ and $v_2$ at random from $C2V\left[c_1\right]$ and $C2V\left[c_2\right]$
\State $E_X = E_X \cup \left\{\left(v_1,v_2\right)\right\}$
\EndFor
\State \Return{$\tup{V_X,E_X,L,L(V_X)}$}
\EndProcedure
\end{algorithmic}
\caption{The input to the algorithm is the source property graph dataset $D_S$ and the number of vertices and edges in the target property graph - $n_t = |V_T|$ and $m_t = |E_T|$. The output is the target property graph $\tup{V_T,E_T,L,L(V_T)}$. }
\label{algo-pgmbasic}
\end{algorithm}

Lines 6 and 7  compute the label and edge connectivity distributions $P_L$ and $P_C$ from the source graph dataset $G_S$. Lines 9-11 sample from the distribution $P_L$, the vertex label set $L(V_T)$ for the target graph. Lines 14-17 construct the edge set $E_T$ by drawing one edge at a time by sampling from a multinomial distribution over the edge categories. The resultant vertices to be connected are drawn at random from the sets of vertices corresponding to the label categories obtained from the edge category. Self and repeated edges can be removed by post-processing.

\section{When Labels Fall Short}
\label{sec:results}
We use two example graphs with contrasting properties to illustrate the strengths and limitations of the PGM method. The first example is a role-based graph \cite{henderson2012rolx} such as an enterprise network where the connectivity depends on roles that the vertices serve \cite{chen2016multi}. Thus, it is possible that there is a high chance of an edge between a SERVER-CLIENT pair while the chance of an edge between a SERVER-SERVER or a CLIENT-CLIENT pair is small. By considering two binary labels that can explain the edge connectivity, we synthesized a role-based graph with 2000 vertices and 90,000 undirected edges. Our next example consists of an anonymized Facebook social graph from the SNAP website \cite{snapnets}. The data is available as a number of ego-nets, each associated with a large number of binary vertex features that vary across the ego-nets. We collected the 4 labels that were common to all vertices across the ego-nets and leveraged the combined graph for our experiments. The graph has around 4000 vertices and 88,000 undirected edges with nearly a power-law degree distribution. 

We then run the steps presented in Algorithm \ref{algo-pgmbasic} to re-generate target property graphs of same size as the source property graphs. We compare the distributions $P_L$ and $P_C$. The design of the algorithm ensures that $P_L$ and $P_C$ for the source and target distributions will match well in expectation and the same was verified. We also quantify the comparison with respect to the degree distributions between the ground truth graph and the regenerated graph by means of the Jenson-Shannon Divergence (JSD) measure \cite{lin1991divergence}. JSD is small when the distributions are closer to each other. 

\begin{figure}[!htbp]
\begin{centering}
\includegraphics[width=0.85\columnwidth]{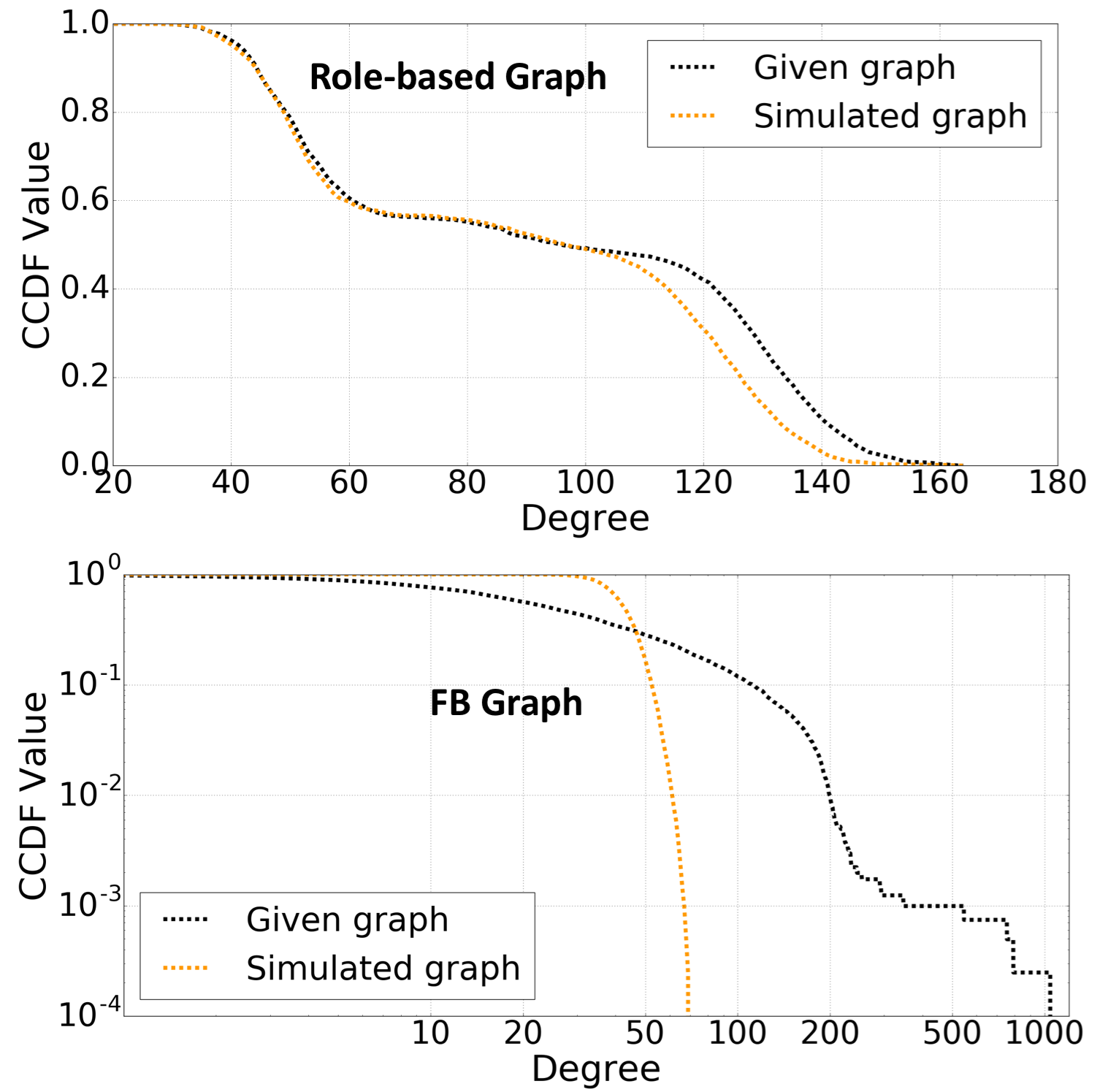}
\caption{\textit{Top: }Degree distribution comparison (linear scale) for the scenario where the graph structure is fully explained by the given labels. (JSD = 0.036). \textit{Bottom: } Degree distribution comparison (log-log scale) for the Facebook graph where the graph structure is not fully explained by the given labels (JSD = 0.354)}
\vspace{-0.6em}
\label{fig:regen}
\end{centering}
\end{figure}

The results for both the example datasets are shown in Figure \ref{fig:regen}. The top sub-figure shows the degree distribution comparisons between the source and regenerated versions of the role-based graph, for which there is a very good match. The degree distribution is plotted as a complementary cumulative distribution function (CCDF).  The bottom sub-figure shows comparisons for the Facebook graph (on a log-log scale) for which we don't see a good match.  

The joint distribution based approach that we described in Algorithm \ref{algo-pgmbasic} assumes that the edge connectivity is a function of label values alone.This assumption is often violated in the case of real-world datasets rendering the basic PGM approach ineffective in recovering the structural properties such as the degree distributions. It might be impossible to identify and collect all the vertex labels that can explain the graph structure. Even if all the possible labels can be collected, it is possible that the graph is grown temporally and as a result, when new vertices arrive and form edges, the connectivity is not only dependent on the label combination pairs but also on the structure of the graph itself at the time point of their arrival. The next section bridges this gap and extends the PGM approach to replicate the topological features under limited label information.

\section{Label Augmentation to Rescue}
\label{sec:extended}

In \cite{papadopoulos2012popularity}, the authors introduce the notion that the probability of edge formation between a new vertex and a vertex already present in the graph is dependent on  both the \textit{similarity} between the two vertices and the \textit{popularity} of already present vertex. The similarity notion refers to affinity based on vertex attributes. The popularity notion captures phenomena such as preferential attachment where vertices tend to get attached to popular vertices which are vertices with existing high degree values. Strict role-based networks such as communication networks will favor similarity while networks such as social networks will favor a combination of similarity and popularity. In the case of the PGM approach, the joint distribution implicitly encodes and generalizes the notion of similarity by quantifying the average likelihood of edge connectivity between all possible pairs of label categories (not necessarily between vertices having the same label categories). The label augmentation process that we describe next, will bring in the popularity aspect into the PGM approach.

Adopting the above philosophy, in order to better match the degree distribution of the given graph, we propose to augment the given set of labels with an additional label $L_a$ that describe the vertex popularity. The number of values that this additional label can take on is denoted by $n_a$, corresponding to the division of the range of the degree values for the given graph $G_S$ into $n_a$ intervals. Vertex-specific label values for  are assigned based on the interval in which a given vertex degree falls. We then run an iterative procedure by incrementing $n_a$ by $1$ at each step till an error measure over the source and target distributions of structural properties of interest is acceptable. In each iteration, the interval lengths can be optimally adjusted by means of an optimizer to minimize the error metric. Note that both the distributions $P_L$ and $P_C$ without $L_a$ will be retained as before due to the marginalization property of the joint probability mass functions. Algorithm \ref{algo-pgm-aug} reflects the updated flow.  
\begin{algorithm}[!htbp]
\begin{algorithmic}[1]
\State $\tup{V_S,E_S,L,L(V_S)}$ = $processDataSet\left(D_S\right)$
\State $n_a \gets 1$ \Comment{$n_a$ is the number of intervals in degree range}
\State $error \gets \infty$
\Procedure{PGM-AUGMENTED}{$\tup{V_S,E_S,L,L(V_S)},n_t,m_t$}
\While {$\left(error > tolerance\right)$}
\State $n_a \gets n_a + 1$
\State Divide degree range of the source graph into $n_a$ intervals
\ForEach {$v \in V_S$}
\State Assign $l_a(v)$ value based on the degree($v$)
\State Append the label vector $\bar{L}(v)$ with $l_a(v)$
\EndFor
\State $G_T$ = \Call{simAttrGraph}{$\tup{V_S,E_S,L,L(V_S)},n_t,m_t$}
\State $error \gets computeError\left(G_S,G_T\right)$
\EndWhile
\EndProcedure
\end{algorithmic}
\caption{The updated approach that uses label augmentation. This algorithm calls the simAttrGraph procedure in Algorithm. \ref{algo-pgmbasic}.}
\label{algo-pgm-aug}
\end{algorithm}
\vspace{-15pt}
\begin{figure}[!htbp]
\begin{centering}

\includegraphics[width=0.85\columnwidth]{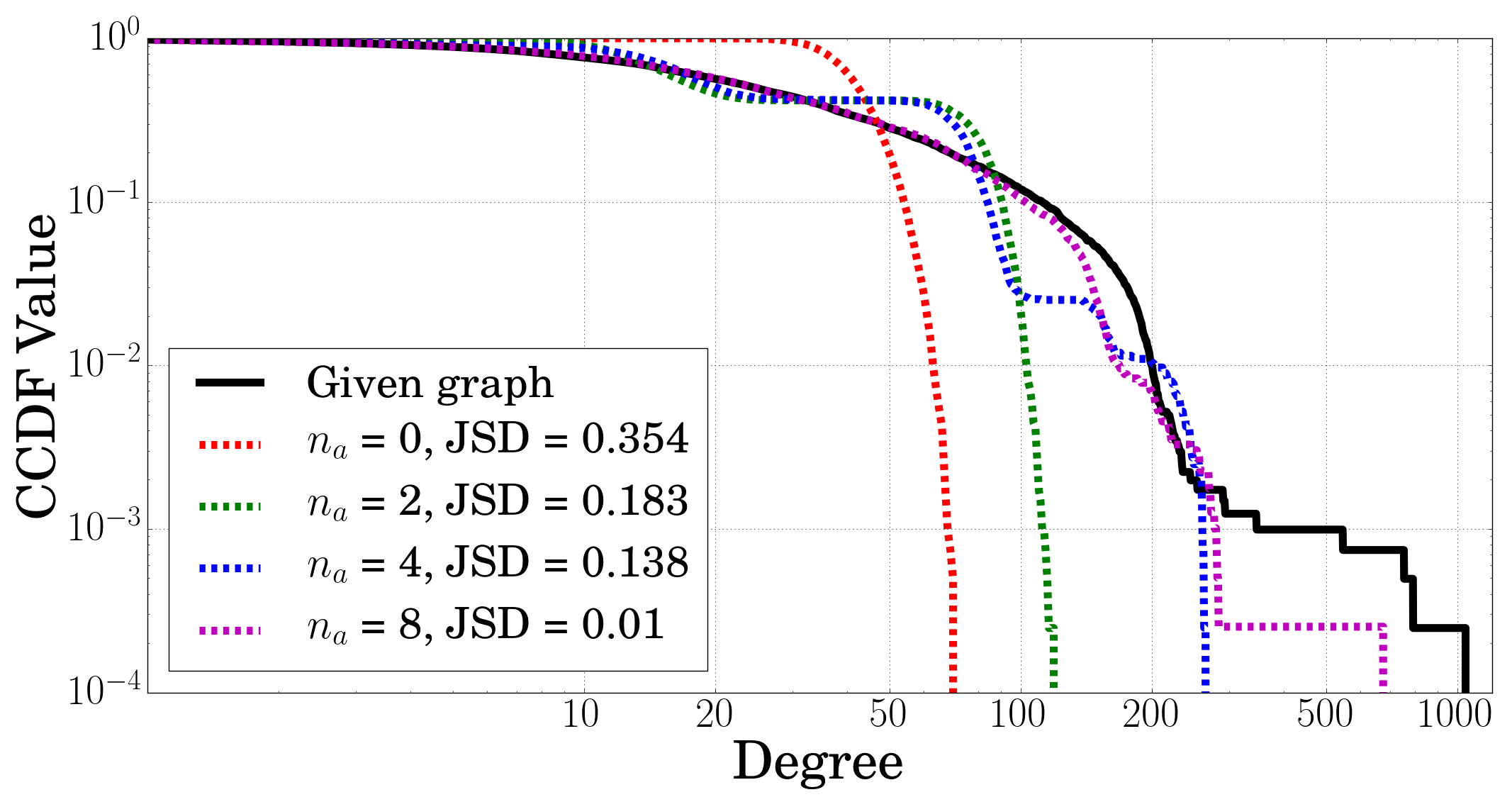}
\caption{Degree distribution comparison (log-log) between the Facebook graph and the simulated graph with augmented label $L_a$ and for various values of $n_a$. }
\vspace{-0.6em}
\label{fig:all_aug}
\end{centering}
\end{figure}

In our experiments, for a given value of $n_a$, we assigned the interval lengths based on a logarithmic scale and the end-points of the intervals were fixed.  For the Facebook graph, the results of  augmenting with $L_a$ with $n_a$ = 0,2,4,8 are shown in Figure \ref{fig:all_aug}. As seen the reproduction of the degree distribution is very poor without augmentation ($n_a=0$) and gets progressively better with augmentation and by increasing $n_a$. The same is reflected in the observation that the JSD measure decreases with increasing $n_a$. 

\section{Dataset expansion}
\label{sec:expand}
\begin{figure}[!htbp]
\begin{centering}
\includegraphics[width=0.85\columnwidth]{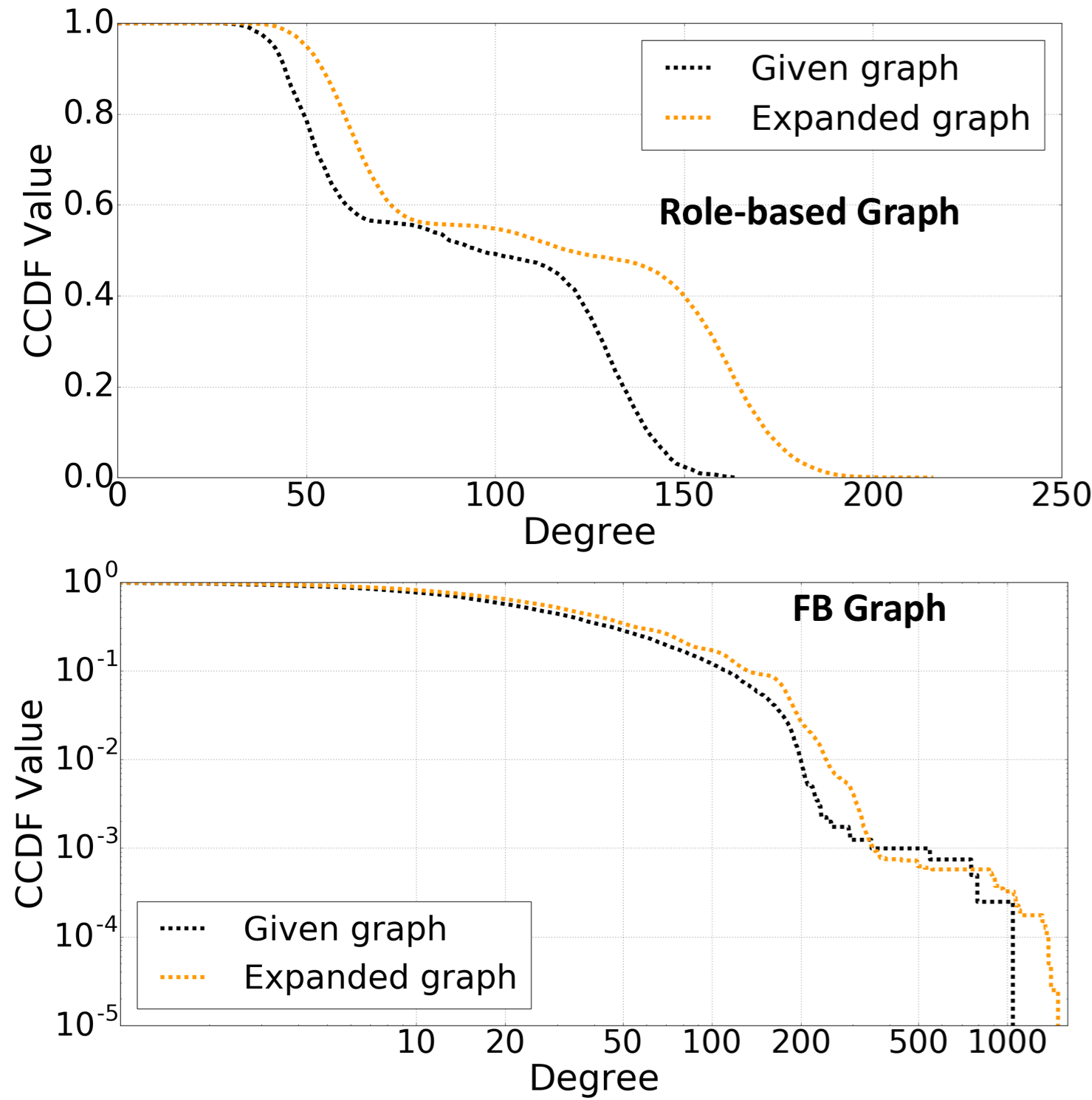}
\caption{Comparing degree distributions shapes for a 10X dataset expansion. \textit{Top :} Role-based graph. \textit{Bottom : } Facebook graph with 8 label values for the augmented label (log-log scale).}
\label{fig:expand}
\end{centering}
\end{figure}

Next we consider the expansion of the dataset by using the estimated joint distributions of the vertex labels and the edge connectivity, $P_L$ and $P_C$ respectively. The results are illustrated in Figure. \ref{fig:expand} for both the role-based and the Facebook graphs. The number of vertices were expanded by 10X where as the number of edges were expanded by 12.5X. It's clear from the observed results that the PGM approach in its basic or extended form works well in expanding the attributed graphs and preserves the degree distribution shapes. Leveraging a serial implementation, we generated graphs with 1 million vertices, 31 million edges and  total of 16 vertex label categories (2 binary labels and an augmented label with 4 values) in about 42 minutes on a 2.6GHz Mac workstation. Drawing of independent edges facilitates easy parallelization of the code which is ongoing. 
\vspace{-8pt}

\section{Related Work}
\label{sec:related}
Synthetic generation of property graphs is a nascent area of research when compared to models for network topologies. Approaches based on exponential random graph (ERG) \cite{robins2007introduction} model the link probability as a linear model in a number of topological features.  While such formulations are general enough to accommodate vertex labels, these methods have high computational cost beyond a few thousand vertices \cite{pfeiffer2014attributed}. The Multiplicative Attribute Graph (MAG) approach models the link probability between two vertices in terms of affinities along a number of independent vertex level latent labels. However, MAG's drawback also lies in its reliance on latent labels. Generating vertex labels as observed in the data becomes difficult in a latent label based approach \cite{pfeiffer2014attributed}. An alternate approach is presented in Attributed Graph Model (AGM) \cite{pfeiffer2014attributed} that combines two sources of information: a) it learns the correlation between vertex labels and the graph structure, and b) exploits a known generative model for the graph topology in the form or Kronecker Product Graph Model or the Chung-Lu model.  The AGM approach can perform well in replicating the graph topology and the correlation with the label values for any given set of labels.  However the approach is agnostic to the explanatory power of the labels. Further, modeling and expanding arbitrary property graph datasets can be a challenge with the AGM approach that relies on specific models for the graph topology.  In a recent work \cite{Sukthankar-SocialInfo2014} the authors focus on the problem of cloning social networks in a privacy preserving fashion.  The authors use preferential attachment model to generate the graph, followed by genetic algorithms to align the statistical distribution of attributes in the source and derived dataset.  The use of optimization process in conjunction with the preferential attachment based model limits the applicability and scalability of this approach.

\section{Conclusions and Ongoing Work}
\label{sec:conclu}
We present a property graph generation algorithm that bridges two state of the art approaches, \cite{kim2012multiplicative} and \cite{pfeiffer2014attributed}, by leveraging on their strengths, and addresses their respective weaknesses in modeling realistic property graphs. We initiate the simulation with observed labels and then introduce an augmented label to explain when the connectivity is not explained by the given set of labels.  Our approach reproduces or expands property graphs with a single edge relation and homogeneous vertices while being able to match the degree distributions closely. We are addressing several theoretical and implementation challenges as part of ongoing research. They include supporting heterogeneous vertices and relationships, better label augmentation strategies for large-scale dataset expansion and preservation of properties beyond degree distribution.

\begin{acks}
This research was supported by the High Performance Data Analytics program at the Pacific Northwest National Laboratory (PNNL). PNNL is a multi- program national laboratory operated by Battelle Memorial Institute for the US Department of Energy under DE-AC06- 76RLO 1830.
\end{acks}

\bibliographystyle{ACM-Reference-Format}

\end{document}